\documentclass[12pt]{article}

\input{tcilatex}

\begin{document}

\begin{abstract}
The example of nonpositive trace-class Hermitian operator for which
Robertson--Schr\"odinger uncertainty relation is fulfilled is presented. The
partial scaling criterion of separability of multimode continuous variable
system is discussed in the context of using nonpositive maps of density
matrices.
\end{abstract}

\begin{center}
{\Large \textbf{DOES THE UNCERTAINTY RELATION DETERMINE THE QUANTUM STATE?}}

\bigskip 
{Olga V. Man'ko,$^1$ V.I. Man'ko,$^1$ G. Marmo,$^2$ E.C.G. Sudarshan$^3$ and
F. Zaccaria$^2$}

{$^1$P.N. Lebedev Physical Institute, Leninskii Prospect, 53, Moscow 119991
Russia\\[0pt]
$^2$Dipartimento di Scienze Fisiche, Universit\`{a} ``Federico II'' di
Napoli and Istituto Nazionale di Fisica Nucleare, Sezione di Napoli,
Complesso Universitario di Monte Sant Angelo, Via Cintia, I-80126 Napoli,
Italy\\[0pt]
$^3$Physics Department, Center for Particle Physics, University of Texas,
78712 Austin, Texas, USA\\[0pt]
}
\end{center}

Emails: {\small \textbf{%
omanko@sci.lebedev.ru~~manko@sci.lebedev.ru~~marmo@na.infn.it\newline
sudarshan@physics.utexas.edu~~zaccaria@na.infn.it }} 

\section{Introduction}

The main difference of quantum and classical mechanics is connected with the
uncertainty relations by Heisenberg~\cite{Heisenberg27} and by Schr\"odinger~
\cite{Schroedinger30} and Robertson~\cite{Robertson30}. The uncertainty
relations were studied in \cite{DodKurmPLA} and \cite{SudarPRA}. A review of
the uncertainty relations is presented in \cite{183}. The uncertainty
relation is used to formulate the separability criterion for composed system
states. For continuous variables, the separability criterion based on
Peres--Horodecki partial transpose of subsystem density matrix~\cite
{PeresPRL,HorodPLA} was applied in \cite{Simon} to the case of two-mode
system using the Schr\"odinger--Robertson uncertainty relation. The partial
scaling transform was suggested as a separability criterion for multimode
continuous-variable systems in \cite{SudOlgaPLA} and the key role is in
using the multimode Robertson uncertainty relation. A recent review is
presented in \cite{Manc-SeverQuant-ph}.

In view of these applications, as well as due to the necessity of a deeper
understanding the role of uncertainty relations in characterizing the set of
all admissible quantum states. In this paper, we address in this paper to
the following question.

Does the uncertainty-relation imply that the density state~\cite
{Landau,vonNeum} used to evaluate it is really Hermitian, trace-class and
nonnegative? In other words, is the uncertainty-relation both a necessary
and sufficient condition for the density state to be nonnegative?

The aim of this paper is to show that the uncertainty-relation fulfilling is
necessary but not sufficient condition for the nonnegativity of the density
matrix. We present the example of the "pseudodensity operator" for which the
uncertainty relation is fulfilled but the operator itself has negative
eigenvalues.

The other goal of this paper is to discuss a possible role of nonpositive
maps to detect the entanglement. We consider the two-mode and multimode
Gaussian states applying the momentum scaling transform which turns out to
be nonpositive.~\footnote{%
E.V. Shchukin directed our attention to the nonpositivity of this map.} We
show how nonpositive map of subsystem density matrix preserving the
uncertainty relation can be used to formulate the separability criterion.

The paper is organized as follows.

In Section 2, we review the standard derivation of the
Schr\"odinger--Robertson uncertainty relation. In Section 3, we describe the
nonpositive map of density operator for continuous variables which is
scaling transform of momentum in the Wigner-function representation of the
quantum state. In Section 4, the example of Hermitian nonpositive operator
with trace equal to unity for which the Schr\"odinger--Robertson uncertainty
relation takes place is presented. In Section 5, we formulate a criterion of
separability using nonpositive scaling map (and the other maps). The
perspectives and conclusions are done in Section 6.

\section{Schr\"odinger--Robertson uncertainty relation}

In this section, we review the derivation of position--momentum uncertainty
relation.

Given nonegative Hermitian operator $\hat\rho$ which is the density operator
of a quantum state with $\mbox{Tr}\,\hat\rho=1$. The obvious relation holds 
\begin{equation}  \label{UR1}
\mbox{Tr}\,(\hat F^\dagger\hat F\hat \rho)\geq 0,
\end{equation}
where $\hat F$ is an arbitrary operator.

Let us take operator $\hat F$ in the form 
\begin{equation}  \label{UR2}
\hat F=c_1\hat q+c_2\hat p,
\end{equation}
where $\hat q$ and $\hat p$ are canonical position and momentum operators,
e.g., for harmonic oscillator, and $c_1$ and $c_2$ are complex numbers.

We rewrite (\ref{UR2}) in the form 
\begin{equation}  \label{UR3}
\hat F=\sum_{\alpha=1}^2c_\alpha\hat Q_\alpha, \qquad\hat Q_1=\hat q,\qquad
\hat Q_2=p_2.
\end{equation}
Inequality (\ref{UR1}) takes the form 
\begin{equation}  \label{UR4}
\sum_{\alpha,\beta=1}^2c_\alpha^*c_\beta\langle\hat Q_\alpha\hat
Q_\beta\rangle\geq 0,\qquad\langle \hat Q_\alpha\hat Q_\beta\rangle=\mbox{Tr}%
\,\hat\rho\hat Q_\alpha\hat Q_\beta.
\end{equation}
In view of the identity 
\begin{equation}  \label{UR5}
\langle\hat Q_\alpha\hat Q_\beta\rangle=\frac12\,\langle\{\hat Q_\alpha,\hat
Q_\beta\}\rangle+\frac12\,\langle[\hat Q_\alpha,\hat Q_\beta]\rangle,
\end{equation}
where we use, as usual, the symmetrized and commutator product, one can
rewrite the positivity condition for quadratic form (\ref{UR4}) in variables 
$c_\alpha$ as the positivity condition of the matrix of quadratic form,
i.e., 
\begin{equation}  \label{UR6}
\langle\hat Q_\alpha\hat Q_\beta\rangle\geq 0\quad\mbox{or}%
\quad\left(\frac12\,\langle\{\hat Q_\alpha,\hat Q_\beta\}\rangle+\frac12\,%
\langle[\hat Q_\alpha,\hat Q_\beta]\rangle\right)\geq 0.
\end{equation}
For the particular operator $\hat F$ given by (\ref{UR2}), the above
condition is the condition of positivity of the Hermitian matrix 
\begin{equation}  \label{UR7}
\left(
\begin{array}{clcr}
\langle\hat q^2\rangle & \frac12\langle\left(\hat q\hat p+\hat p\hat q
\right)\rangle+\frac{i}{2} &  &  \\ 
\frac12\langle\left(\hat q\hat p+\hat p\hat q \right)\rangle-\frac{i}{2} & 
\langle\hat p^2\rangle &  & 
\end{array}
\right)\geq 0.
\end{equation}

In view of the Sylvester criterion for the positivity of the matrix, one has
obvious inequalities 
\[
\langle\hat q^2\rangle\geq 0,\qquad \langle\hat p^2\rangle\geq 0
\]
and the uncertainty relation 
\begin{equation}  \label{UR8}
\langle\hat q^2\rangle\langle\hat p^2\rangle-\frac14\langle\left(\hat q\hat
p+\hat p\hat q \right)\rangle^2\geq\frac14\,,\quad \hbar=1.
\end{equation}
Now if one replaces in (\ref{UR2}) $\hat q\rightarrow\hat q-\langle\hat
q\rangle$, $\hat p\rightarrow\hat p-\langle\hat p\rangle$, relation (\ref
{UR8}) is the Schr\"odinger--Robertson uncertainty relation 
\begin{equation}  \label{UR9}
\sigma_{qq}\sigma_{pp}-\sigma_{qp}^2\geq \frac14\,,
\end{equation}
where the variances of position and momentum 
\begin{equation}  \label{UR10}
\sigma_{qq}=\langle\hat q^2\rangle-\langle\hat q\rangle^2,\qquad
\sigma_{pp}=\langle\hat p^2\rangle-\langle\hat p\rangle^2
\end{equation}
and covariance of position and momentum 
\begin{equation}  \label{UR11}
\sigma_{qp}=\frac12\,\langle\left(\hat q\hat p+\hat p\hat q
\right)\rangle-\langle\hat q\rangle\langle\hat p\rangle
\end{equation}
are involved.

In the case of $\langle\hat q\rangle=0$ and $\langle\hat p\rangle=0$,
inequalities (\ref{UR9}) and (\ref{UR8}) are equivalent.

For the multimode case, one takes the operator 
\begin{equation}  \label{UR12}
\hat F=\sum_{\alpha=1}^{2N}c_\alpha\hat Q_\alpha,\quad \hat Q_1=\hat q_1,
\hat Q_2=\hat q_2,\ldots,\hat Q_N=\hat q_N,\hat Q_{N+1}=\hat p_1,\ldots,\hat
Q_{2N}=\hat p_N.
\end{equation}
In this case, relation (\ref{UR1}) provides the positivity condition for
matrix (\ref{UR5}) where $\alpha,\beta=1,\ldots, 2N$. This condition can be
rewritten as the positivity of the matrix principal minors. The determinant
of the matrix (last $2N$th principal minor) yields the weaker inequality 
\begin{equation}  \label{UR13}
\mbox{det}\left(\frac12\langle\hat Q_\alpha\hat Q_\beta+\hat Q_\beta \hat
Q_\alpha\rangle\right)\geq\frac{1}{4^N}.
\end{equation}

\section{Scaling transform as nonpositive map}

There are completely positive and not completely positive linear maps of
density operators~\cite{Sudar}. Usually for detecting the entanglement one
uses positive but not completely positive maps. In \cite{SudOlgaPLA} we used
the scaling transform to study the separability criterion for continuous
variables.

Below we show that the scaling transform is a nonpositive transform. The
scaling transform is defined through the transform of Wigner function 
\begin{equation}  \label{ST1}
W(q,p)\rightarrow W_s(q,p)=|\lambda|^2W(\lambda q,\lambda p).
\end{equation}
The Wigner function is related to density matrix $\rho(x,x^{\prime})$ in the
position representation by the invertible map 
\begin{equation}  \label{ST2}
W(q,p)=\int\rho\left(q +\frac{u}{2}\,,q-\frac{u}{2}\right) e^{-ip u}\,du
\end{equation}
and 
\begin{equation}  \label{ST3}
\rho(x,x^{\prime})=\frac{1}{2\pi}\int W\left(\frac{x+x^{\prime}}{2}%
\,,p\right) e^{ip(x-x^{\prime})}\,dp.
\end{equation}
For the case of $\mbox{Tr}\rho=1$, one has the normalization condition 
\begin{equation}  \label{ST4}
\int W(q,p)\frac{dq\,dp}{2\pi}=1.
\end{equation}
One can see that if $W(q,p)$ satisfies the condition (\ref{ST4}), due to the
factor $|\lambda|^2$ in (\ref{ST1}), we obtain 
\begin{equation}  \label{ST5}
\int W_s(q,p)\frac{dq\,dp}{2\pi}=1.
\end{equation}
One can use the squeezing transform 
\begin{equation}  \label{ST6}
W(q,p)\rightarrow W^{\mathrm{sq}}(q,p)=W(\kappa q,\kappa^{-1}p).
\end{equation}
The squeezing transform is a unitary transform. Due to this, the nonnegative
density operator is mapped by the squeezing transform onto another
nonnegative density operator. The combination of transform (\ref{ST1}) and
squeezing transform provides the map of Wigner function used in \cite
{SudOlgaPLA} 
\begin{equation}  \label{ST7}
W(q,p)\rightarrow W^{\mathrm{ps}}(q,p)=|\lambda|W(q,\lambda p).
\end{equation}
In fact, taking $\kappa^2=\lambda^{-1}$ in (\ref{ST6}) and making then
transform (\ref{ST1}) with scaling parameter $\sqrt\lambda$, we get (\ref
{ST7}).

Below we show that the transform (\ref{ST1}) is also nonpositive. This means
that (\ref{ST7}) is also nonpositive since it is the product of nonpositive
scaling transform and positive squeezing transform (\ref{ST1}).

To show that the scaling transform is nonpositive, we calculate the fidelity 
\begin{equation}  \label{ST8}
f=\mbox{Tr}\,\rho_0\rho_1^{(s)}=\langle 0\mid\rho_1^{(s)}\mid0\rangle,
\end{equation}
where $\rho_0$ is density operator $\mid 0\rangle\langle 0\mid$ of ground
state $\mid 0\rangle$ of harmonic oscillator and $\rho_1^{(s)}$ is the
scaled density operator of the first excited state of the harmonic
oscillator (we take $\hbar=m=\omega=1)$. The fidelity can be calculated in
terms of overlap integral of Wigner functions of the corresponding states.

The Wigner function of the ground state reads (see, e.g., \cite{183}) 
\begin{equation}  \label{ST9}
W_0(q,p)=2e^{-q^2-p^2}.
\end{equation}

The Wigner function of the first excited state reads 
\begin{equation}  \label{ST10}
W_1(q,p)=2(2q^2+2p^2-1)e^{-q^2-p^2}.
\end{equation}

The scaled Wigner function depends on the parameter $\lambda$, i.e., 
\begin{equation}  \label{ST11}
W_1^{(s)}(q,p)=2|\lambda|^2[2\lambda^2(q^2+p^2)-1]e^{-\lambda^2q^2
-\lambda^2p^2}.
\end{equation}
For small parameter $\lambda$, the leading term in (\ref{ST1}) reads 
\begin{equation}  \label{ST12}
W_1^{(s)}(q,p)\approx -2|\lambda|^2.
\end{equation}
In view of this, the fidelity is 
\begin{equation}  \label{ST13}
f=\int W_0(q,p)W_1^{(s)}(q,p)\frac{dq\,dp}{2\pi}\approx -2|\lambda|^2<0.
\end{equation}
This means that at least one diagonal matrix element of scaled density
operator is negative. Thus, for small parameters of scaling, the map under
discussion is nonpositive.

Though we consider scaling map it is worth noting that there exists another
map of density operators providing smoothed Wigner function from the initial
one. For this map, one uses as convolution kernel some other Wigner function
as it was shown in \cite{sm87} this map is nonpositive for some choices of
the Wigner function kernel. In particular, the convolution kernel based on
Wigner function of the first excited state of the harmonic oscillator is of
this kind.

\section{Wigner--Weyl symbol of nonpositive Hermittian operator with
fulfilling the uncertainty relation}

In this section, we use the property of nonpositive map to present the
Hermitian nonpositive operator $\hat\rho$ with $\mbox{Tr}\,\hat\rho=1$ $%
(\hat\rho<0)$ fulfilling the condition 
\begin{equation}  \label{ST14}
\left(
\begin{array}{clcr}
\mbox{Tr}\,\hat\rho\hat q^2 & \mbox{Tr}\,\hat\rho\frac12\left(\hat q\hat
p+\hat p\hat q \right) &  &  \\ 
\mbox{Tr}\,\hat\rho\frac12\left(\hat q\hat p+\hat p\hat q \right) & \mbox{Tr}%
\,\hat\rho\hat p^2 &  & 
\end{array}
\right)> 0.
\end{equation}
In fact, this operator has matrix elements in the position representation
determined by scaled Wigner function (\ref{ST11}), i.e., 
\begin{equation}  \label{ST15}
\rho^{(s)}(x,x^{\prime})=\frac{\lambda^2}{2\pi}\int 2 \left[2\lambda^2\left(%
\Big(\frac{x+x^{\prime}}{2}\Big)^2+p^2\right)-1\right] e^{-\lambda^2[(x+x^{%
\prime})/2]^2-\lambda^2p^2}e^{ip(x-x^{\prime})}\,dp.
\end{equation}
This "density" operator is negative but the variances and covariance of
position and momentum read $(\langle q\rangle=0$, $\langle p\rangle=0)$ 
\begin{eqnarray}  \label{ST16}
&&\sigma_{qq}=\langle\hat q^2\rangle=\frac{1}{\lambda^2}\sigma_{qq}^{(1)}=%
\frac{3}{2\lambda^2},  \nonumber \\
&&\sigma_{qp}=\frac12\left(\hat q\hat p+\hat p\hat q\right)=0, \\
&&\sigma_{pp}=\langle\hat p^2\rangle=\frac{1}{\lambda^2}\sigma_{pp}^{(1)}=%
\frac{3}{2\lambda^2}.  \nonumber
\end{eqnarray}
Here $\sigma_{qq}^{(1)}$ and $\sigma_{pp}{(1)}$ are quadrature dispersions
in the first excited states of the harmonic oscillator.

The uncertainty relation is respected for $|\lambda|\leq 1$. Thus, for small
scaling parameters, the uncertainty relation is fulfilled. Nevertheless, the
"density" operator used to calculate the dispersion matrix is nonpositive.
Thus we demonstrated that fulfillment of the Schr\"odinger--Robertson
uncertainty relation does not mean that the operator used as density
operator must be positive. There can exist negative operators, for which
matrix (\ref{UR7}) calculated as if the operators $\hat\rho$ are density
operators, is positive. Thus, the set of operators fulfilling the
uncertainty relations in the form of positivity of matrix (\ref{UR7}) is
broader than the set of density states. To obtain the set of these
operators, one needs to solve the inverse problem formulated as follows.

Given matrix (\ref{UR7}).

What are the operators which provide nonnegativity of this matrix?

The symplectic transform of positive density operators keeps the operators
positive. The positivity of matrix (\ref{UR7}) is invariant property with
respect to symplectic transform. But there exist extra transforms belonging
to general linear group which preserve the positivity of the matrix \cite
{Grab}. Scaling is just one of such transforms. Being expressed in terms of
a map of density operators, these transforms provide the negative map
respecting the uncertainty relations.

\section{Separability criterion and nonpositive maps}

The separable state of two-mode system is described by the density operator $%
\hat\rho(1,2)$ which can be represented as convex sum of simply separable
states, i.e., 
\begin{equation}  \label{ST17}
\hat\rho(1,2)=\sum_kp_k\hat\rho^{(k)}(1)\otimes\widetilde{\hat\rho^{(k)}}%
(2), \qquad p_k\geq 0.
\end{equation}
The Wigner function of the separable state has the form 
\begin{equation}  \label{ST18}
W(q_1,p_1,q_2,p_2)=\sum_kp_kW^{(k)}(q_1,p_1)\widetilde W^{(k)}(q_2,p_2).
\end{equation}
If one takes nonpositive partial scaling transform of the subsystem states 
\[
\widetilde W^{(k)}(q_2,p_2)\rightarrow |\lambda|\widetilde
W^{(k)}(q_2,\lambda p_2)
\]
suggested in \cite{SudOlgaPLA}, the uncertainty relation is respected for $%
|\lambda|\leq 1$ which means that the initial nonnegative matrix 
\begin{equation}  \label{ST19}
\|A_{ij}\|=\left(
\begin{array}{clcr}
\sigma_{q_1q_1} & \sigma_{q_1p_1}+\frac{i}{2} & \sigma_{q_1q_2} & 
\sigma_{q_1p_2} \\ 
\sigma_{p_1q_1}-\frac{i}{2} & \sigma_{p_1p_1} & \sigma_{p_1q_2} & 
\sigma_{p_1p_2} \\ 
\sigma_{q_2q_1} & \sigma_{q_2p_1} & \sigma_{q_2q_2} & \sigma_{q_2p_2}+\frac{i%
}{2} \\ 
\sigma_{p_2q_1} & \sigma_{p_2p_1} & \sigma_{p_2q_2}-\frac{i}{2} & 
\sigma_{p_2p_2}
\end{array}
\right)\geq 0
\end{equation}
after the scaling transform takes the form 
\begin{equation}  \label{ST20}
\|A_{ij}^{(s)}\|=\left(
\begin{array}{clcr}
\sigma_{q_1q_1} & \sigma_{q_1p_1}+\frac{i}{2} & \sigma_{q_1q_2} & 
\lambda^{-1}\sigma_{q_1q_2} \\ 
\sigma_{p_1q_1}-\frac{i}{2} & \sigma_{p_1p_1} & \sigma_{p_1q_2} & 
\lambda^{-1}\sigma_{p_1p_2} \\ 
\sigma_{q_2q_1} & \sigma_{q_2p_1} & \sigma_{q_2q_2} & \lambda^{-1}%
\sigma_{q_2p_2}+\frac{i}{2} \\ 
\lambda^{-1}\sigma_{p_2q_1} & \lambda^{-1}\sigma_{p_2p_1} & 
\lambda^{-1}\sigma_{p_2q_2} -\frac{i}{2} & \lambda^{-1}\sigma_{p_2p_2}
\end{array}
\right)\geq 0,
\end{equation}
and one must have for separable states 
\begin{equation}  \label{ST21}
\|A_{ij}^{(s)}\|\geq 0.
\end{equation}
Inequality (\ref{ST21}) follows from the condition that each matrix 
\begin{equation}  \label{ST22}
\|A_{ij}^{(s)}\|^{(k)}\geq 0
\end{equation}
and for separable states the convex sum of nonnegative matrices is
nonnegative, i.e., 
\begin{equation}  \label{ST23}
\|A_{ij}^{(s)}\|=\sum_kp_k\|A_{ij}^{(s)}\|^{(k)}\geq 0.
\end{equation}
In our consideration, we employ the condition $\langle\hat{\vec q}\rangle=0$%
, $\langle\hat{\vec p}\rangle=0$ but this condition can be removed. In fact,
if in the initial state $\langle\hat{\vec q}\rangle\neq 0$, $\langle\hat{%
\vec p}\rangle\neq 0$, one can make local unitary shift transform which do
not affect the entanglement properties. For new shifted density operators,
one can apply the arguments presented above.

On the basis of experience to apply nonpositive scaling map for the
detection of the entanglement, one can formulate general scheme of using
negative maps to study the separability and entanglement.

To do this, one needs to generalize the procedure, in view of the
uncertainty relation. In fact, if one has $N^2$ operators labeled as $\hat
A_{ij}$, $j,k=1,2,\ldots,N$, one may construct the matrix 
\begin{equation}  \label{ST24}
\|A_{ij}\|=\|\mbox{Tr}\left(\hat\rho\hat A_{ij}\right)\|,
\end{equation}
where $\hat\rho$ is any Hermitian operator (not necessarily a nonegative
density operator). Having the set of operators $\hat\rho_k$ one has the set
of matrices 
\begin{equation}  \label{ST25}
\|A_{ij}\|^{(k)}=\|\mbox{Tr}\left(\hat\rho_k\hat A_{ij}\right)\|.
\end{equation}
Assume that $\|A_{ij}\|^{(k)}\geq 0$, then a convex combination is also
nonnegative $\sum_kp_k\|A_{ij}\|^{(k)}\geq 0$. With the help of these
remarks we return to the separability criterion.

Given separable state (\ref{ST17}) of bipartite system, one can apply
nonpositive $\mathcal{N}$ map to the second-subsystem density matrix 
\begin{equation}  \label{ST28}
\widetilde{\hat\rho^{(k)}}(2)\rightarrow\mathcal{N}\widetilde{\hat\rho^{(k)}}%
(2)=\widetilde{\hat\rho_{\mathcal{N}}^{(k)}}(2).
\end{equation}
This operation induces the map 
\[
\hat\rho(1,2)\rightarrow\hat\rho_{\mathcal{N}}(1,2).
\]
The operator $\widetilde{\hat\rho_{\mathcal{N}}^{(k)}}(2)$ can be
nonpositive but we assume extra conditions, namely, there exist the set of
operators $\hat A_{ij}$ for which the numerical matrices 
\begin{eqnarray}
&&\|\mbox{Tr}\,\hat A_{ij}\left(\hat\rho^{(k)}(1)\otimes\widetilde{%
\hat\rho^{(k)}}(2) \right)\|\geq 0,  \label{ST29} \\
&&\|\mbox{Tr}\,\hat A_{ij}\left(\hat\rho^{(k)}(1)\otimes\widetilde{%
\hat\rho^{(k)}}_{\mathcal{N}}(2)\right)\|\geq 0.  \label{ST30}
\end{eqnarray}
Then for separable states the convex sum of nonnegative matrices (\ref{ST30}%
) yields 
\[
\mbox{Tr}\,\|\hat\rho_{\mathcal{N}}(1,2)\hat A_{ij}\|\geq 0.
\]
For entangled states, one can have violation of this inequality. Thus in our
formulation of using nonpositive maps to detect the entanglement we
introduce a new element. It is a set of operators $\hat A_{ij}$ labeled by
matrix indices. After tracing with some positive or negative operator $%
\hat\rho$ the obtained numerical matrices must be positive. This means that
we use the map $\hat\rho\rightarrow\|A_{ij}\|$ of the Hermitian operators
onto positive numerical matrices. This map can be realized in two steps. One
step is the positive or nonpositive map $\hat\rho\rightarrow\hat\rho_{%
\mathcal{N}}$. The second step is $\hat\rho_{\mathcal{N}}\rightarrow\|A_{ij}%
\|=\mbox{Tr}\left(\hat \rho_{\mathcal{N}}\hat A_{ij}\right)$.

Using such procedure one can extend the method of detecting the entanglement
by means of positive but not completely positive maps to apply nonpositive
maps to density operators of composed systems. Namely this ansatz is used
for the partial scaling transform procedure.

\section{Conclusions and Perspectives}

To conclude, we formulate the main results of this study.

We have shown that fulfilling the Schr\"odinger--Robertson
position--momentum uncertainty relation does not imply that the density
operator is nonnegative, i.e., the uncertainty relation does not determine
the quantum state. Fulfilling the uncertainty relation is necessary but not
sufficient condition of nonnegativity of the density operator.

We presented the example of nonpositive operator (in the form of its
Wigner--Weyl symbol) for which the uncertainty relation is fulfilled.

The obtained experience provided the possibility to formulate a procedure of
detecting the entanglement of the multipartite system states using
nonpositive maps of the subsystem density matrices.

It was emphasised that the partial scaling transform criterion suggested in 
\cite{SudOlgaPLA} uses positive map of the position--momentum dispersion
matrix induced by nonpositive map of density operator by means of scaling
momentum in Wigner function.

\section*{Acknowledgments}

O.~V.~M, V.~I.~M. and E.~C.~G.~S. thank Dipartimento di Scienze Fisiche,
Universit\'a ``Federico~II'' di Napoli and Istitito Nazionale di Fisica
Nucleare, Sezione di Napoli for kind hospitality. E.~C.~G.~S. also
acknowledges TAMU Grants No. N00014-04-1-0336 and N00014-03-1-0639.


\begin{thebibliography}{99}
\bibitem{Heisenberg27}  W. Heisenberg, \textsl{Z. Phys.}, \textbf{43}, 172
(1927).

\bibitem{Schroedinger30}  E. Schr\"{o}dinger, \textsl{Sitzungsber. Preuss.
Acad. Wiss.\thinspace }, \textbf{24}, 296 (1930).

\bibitem{Robertson30}  H. P. Robertson, \textsl{Phys. Rev.}, \textbf{35},
667 (1930).

\bibitem{DodKurmPLA}  V.~V.~Dodonov, E.~V.~Kurmushev, and V.~I.~Man'ko, 
\textsl{Phys. Lett.} A, \textbf{79}, 150 (1980).

\bibitem{SudarPRA}  E. C. G. Sudarshan, Charles B. Chiu and G. Bhamathi, 
\textit{Phys. Rev.}, \textbf{52}, 43 (1995).

\bibitem{183}  V.~V.~Dodonov and V.~I.~Man'ko, \textit{Invariants and the
Evolution of Nonstationary Quantum Systems}, \textsl{Proceedings of the
Lebedev Physical Institute}, Nauka, Moscow (1987), Vol.~183 [translated by
Nova Science, New York (1989)].

\bibitem{PeresPRL}  A. Peres, \textit{Phys. Rev. Lett.}, \textbf{77}, 1413
(1996).

\bibitem{HorodPLA}  M. Horodecki, P. Horodecki and R. Horodecki, \textit{%
Phys. Lett.} A, \textbf{223}, 1 (1996).

\bibitem{Simon}  R. Simon 2000 \textit{Phys. Rev. Lett.} \textbf{84} 2726

\bibitem{SudOlgaPLA}  Olga V. Man'ko, V. I. Man'ko, G. Marmo, Anil Shaji, E.
C. G. Sudarshan, F. Zaccaria, \textit{Phys. Lett.} A, \textbf{339}, 194
(2005).

\bibitem{Manc-SeverQuant-ph}  S. Mancini and S. Severini, The Quantum
Separability Problem for Gaussian States, ArXiv cs.CC/0603047.

\bibitem{Landau}  L. D. Landau, \textit{Z. Phys.}, \textbf{45}, 430 (1927).

\bibitem{vonNeum}  J. von Neumann, \textit{Mathematische Grundlagen der
Quantenmechanik}, Springer, Berlin (1932).

\bibitem{Sudar}  E. C. G. Sudarshan, P. M. Mathews and J. Rau, \textit{Phys.
Rev.}, \textbf{121}, 920 (1961).

\bibitem{sm87}  R. Jagannathan, R. Simon, E. C. G. Sudarshan, and R.
Vasudevan, \textit{Phys. Lett.} A, \textbf{120}, 161 (1987).

\bibitem{Grab}  J. Grabowski, M. Kus, and G. Marmo, ArXiv Math-ph 0507045.
\end{thebibliography}
\end{document}